\documentstyle[11pt,aaspp4,epsf]{article}
\newcommand{\be}{\begin{equation}}
\newcommand{\ee}{\end{equation}}
\newcommand{\bea}{\begin{eqnarray}}
\newcommand{\eea}{\end{eqnarray}}
\newcommand{\bc}{\begin{center}}
\newcommand{\ec}{\end{center}}
\newcommand{\etal}{{\it et al.}}
\newcommand{\COBE}{{\it COBE}}

\def\check_mode#1{\ifmmode{#1}\else{$#1$}\fi}

\def\ddeg   {\check_mode{{\rlap.}^\circ}}

\lefthead{Wright, Hinshaw \& Bennett}
\righthead{Differential Map Inversion}

\slugcomment{astro-ph/9510102: Submitted to the Astrophysical Journal Letters}

\begin{document}

\title{Producing Mega-pixel CMB Maps from Differential Radiometer Data}

\author{
E.~L.~Wright\altaffilmark{1},
G.~Hinshaw\altaffilmark{2},
\& C.~L.~Bennett\altaffilmark{3}}

\altaffiltext{1}{UCLA Physics \& Astronomy Dept., Los Angeles CA 90095-1562
(I: wright@astro.ucla.edu)}
\altaffiltext{2}{Hughes STX Corporation, Laboratory for Astronomy and Solar
Physics, Code 685, NASA/GSFC, Greenbelt MD 20771}
\altaffiltext{3}{Laboratory for Astronomy and Solar Physics, Code 685,
NASA/GSFC, Greenbelt MD 20771}

\begin{abstract}
A major goal of cosmology is to obtain sensitive, high resolution maps of the
Cosmic Microwave Background (CMB) anisotropy.  Such  maps, as would be produced
by the recently proposed Microwave Anisotropy Probe ({\em MAP}), will contain a
wealth of primary information about conditions in the early universe.  To
mitigate systematic effects when observing the microwave background, it is
desirable for the raw data to be collected in differential form: as a set of
temperature differences between points in the sky.  However, the production of
large (mega-pixel) maps from a set of temperature differences is a potentially
severe computational challenge.  We present a new technique for producing maps
from differential radiometer data that has a computational cost that grows in
the slowest possible way with increasing angular resolution and number of map
pixels.  The required central processor (CPU) time is proportional to the
number of differential data points and the required random access memory (RAM)
is proportional to the number of map pixels.   We test our technique, and
demonstrate its feasibility, by simulating one year of a space-borne anisotropy
mission.

\end{abstract}



\section{Introduction}

Experiments to measure the anisotropy of the Cosmic Microwave Background
Radiation need to measure signals $\Delta T$ of ${\cal O}(10 \;\mu$K)
in the presence of the large isotropic background with $T_\circ = 2.73$~K
and system noise that is usually ${\cal O}(100\;$K).  The difficulty inherent
in stabilizing the sensitivity of an instrument to much better than 1 part
per million has led all CMB anisotropy experiments to use differential
radiometers, which chop rapidly between the spot whose temperature is to be
measured and one or more reference sources.  The reference sources are usually
other parts of the sky, although occasionally stabilized loads have been used.
The resulting measurements of $\Delta T$ can be analyzed directly in terms of
model predictions, but for high resolution experiments covering a large
fraction of the sky, the number of independent data points can make analysis
extremely cumbersome.  It is far more desirable to reduce the data by
producing a {\it map} of the microwave temperature.  For a given angular
resolution, a map provides the most complete form of data reduction possible
without loss of information, which in turn permits a full range of statistical
tests to be performed on the data.  The problem of converting a large set of
differential sky observations into a map becomes increasingly challenging as
the angular resolution, number of map pixels, and data sampling rate all
increase.  The methods employed by \COBE\ DMR can not be directly adapted to
the high resolution case because of limitations imposed by computer RAM.  We
present a new technique for producing maps which is equivalent to the \COBE\
sparse matrix technique, and which has a computational cost that grows in the
slowest possible way with increasing angular resolution and number of map
pixels: the required CPU time is proportional to the number of differential
data points and the required RAM is proportional to the number of map pixels.
We demonstrate the feasibility of this technique by simulating a one-year
differential mission that produces sky maps with 1,572,864 square pixels of
size 0\ddeg16.

\section{The DMR Experience}

The only full-sky CMB maps produced to date use the differential data from
the \COBE\ DMR.  The DMR data analysis required the construction of sky maps
using the $6 \times 10^7$ differences that were collected each year
in each of the six channels. Since there are only about $10^3$ beam areas on
the sky, the system relating differences to sky temperature is highly
over-determined.  The \COBE\ team chose to analyze the data using
6,144 pixels to cover the sky.  These all have approximately the same area
$\theta_{pix}^2$ and are arranged in a square grid of $32\times 32$ pixels on
each of the 6 faces of a cube.  Within each pixel it was assumed that the
temperature is constant, so the basic problem can be represented as a
least-squares system with $24\times 10^7$ equations (after 4 years of data
taking) in 6,144 variables.  It is also desirable to extend this system to
additionally account for systematic signals in the data due, for example, to
the magnetic sensitivity of the ferrite switches used in the DMR instrument
(Kogut \etal\markcite{SysErr} 1992).  After calibration and baseline
subtraction, one can represent the predicted output of the DMR instrument, in
mK, as
\be
S(t) = V(t)X^T
\ee
where $t$ is the observation time. $V(t)$ is a vector of length $N_{pixels}+3$
given by
\be
V(t) = [0,\ldots,0,+1,0,\ldots,0,-1,0,\ldots,B_x(t),B_y(t),B_z(t)]
\ee
with the +1 in the pixel $p(t)$ that contains the plus horn line-of-sight
at time $t$, the -1 in the pixel $m(t)$ that contains the minus horn
line-of-sight, and $B_i(t)$ being the $ith$ component of the magnetic field in
spacecraft-fixed coordinates.  The parameter vector $X$ is given by
\be
X = [T_0,\ldots,T_{6143},M_x,M_y,M_z]
\ee
where $T_i$ is the temperature of the $i^{th}$ pixel in mK, and
and the $M_i$ are the susceptibilities of the ferrite switches to external
magnetic fields, in mK Gauss$^{-1}$.  For a constant noise per observation
the least squares solution for the map is
(Lineweaver \etal\markcite{COR_NOISE}\ 1994;
Wright \etal\markcite{RMS10} 1994a; Wright \etal\markcite{PSPECT}\ 1994b)
\bea
A  & =  & \sum_t V(t)^T V(t) \nonumber \\
B  & =  & \sum_t S(t) V(t)^T \nonumber \\
X & = & lim_{\epsilon \rightarrow 0^+} \left(A + \epsilon I\right)^{-1}B
\eea
Note that $A$ is {\it sparse}, {\it symmetric}, and {\it singular}.
The main diagonal of $A$ has elements $A_{ii} = N^{obs}_i$, the number of
times each pixel was observed.  The off-diagonal elements of $A$
indicate the number of times a given pixel-pair $ij$ was observed, either
$i = p(t)$ and $j = m(t)$ or vice-versa.  Since the angle between
the DMR horns was fixed at $60^\circ$, only pixel pairs separated
by $60^\circ \pm \theta_{pix}$ could ever be observed.
Thus, while $A$ has $38 \times 10^6$ elements, all but $1.8 \times 10^6$
of these elements are zero, and since $A_{ij} = A_{ji}$, only $9 \times 10^5$
elements need to be kept.  The limit $\epsilon \rightarrow 0^+$ regulates
the singularity of $A$, and gives a map with zero mean.

The matrix $A$ is sparse, but inverting $A+\epsilon I$ does not preserve
this sparseness.  Thus an iterative solution to the system
$(A+\epsilon I)X = B$ is found using algorithms that only require one to
be able to find the product of $A$ and a vector $Y$.
If we let $D$ be a diagonal matrix whose diagonal elements are
$N^{obs}_i$, then a very simple iterative scheme for computing $X$
is
\be
X^{(n+1)} = X^{(n)} + D^{-1}(B-AX^{(n)})
\label{eq.iterate}
\ee
Since differential observations do not constrain the mean of the map,
the mean should be adjusted to zero after each iteration.  With
the DMR scan pattern this simple iteration scheme converges quite
rapidly.  The computational cost of this algorithm for both RAM and
CPU time is proportional to the number of non-zero off-diagonal elements
in $A$.

\section{Scaling for Smaller Beams}

The number of off-diagonal elements can be calculated by
considering the number of pixels in the reference ring of a given pixel.
The radius of the reference ring is the chop angle, $\theta_{chop}$.
The width of the reference ring is about twice the pixel size,
$2 \theta_{pix}$.  The solid angle of the the reference ring is then
$4 \pi \theta_{pix} \sin \theta_{chop}$, and the number of pixels in
the reference ring is $4 \pi \sin \theta_{chop}/\theta_{pix}$.
Since the total number of pixels is $N_{pix} = 4\pi/\theta_{pix}^2$,
the number of non-zero off-diagonal elements in the upper triangle of $A$ is
$N_{off} \approx 8\pi^2 \sin \theta_{chop}/\theta_{pix}^3$.
For the proposed new generation of anisotropy experiments, {\em COBRAS/SAMBA,
FIRE, PSI} and {\em MAP}, the beam sizes will be 10 to 60 times
smaller than the $7^\circ$ beam of the \COBE\ DMR.  Thus the RAM and
CPU time required to invert a differential map by this technique would
increase by a factor of $10^3$ to $2 \times 10^5$ over the time required
for the \COBE\ DMR maps.  Because of this, several groups have abandoned
the wide-angle differential radiometer design that worked so well in the
\COBE\ DMR, and are using either total power radiometers or chopping against
a load, both of which are prone to serious systematic effects.  However, we
have found a different way of executing the operations in Equation
\ref{eq.iterate} that reduces the computational load for small beams
by a substantial factor.

\section{The Time Ordered Alternative}

Many iterative methods for solving $AX = B$ only require the ability to
multiply a vector times $A$.  This multiplication can be performed
by a sum over the elements of $A$, or it can be performed by scanning the
data in time-order and processing each differential pixel pair as it occurs.
The value of an off-diagonal element in $A$ is just the number of times a
given pixel pair occurs, and if the typical off-diagonal element
is large then using the sparse matrix is much faster than scanning the
time-ordered data.  But the typical value of an off-diagonal element scales
like $N_t/N_{off}$, where $N_t$ is the total number of data samples taken.
The proposed new experiments are all at least 10 times more sensitive
per sample than the \COBE\ DMR, so $N_t$ will be only somewhat larger than
the \COBE\ DMR data set, while $N_{off}$ will be at least $10^3$ times larger
than it was for DMR.  Thus for maps with a large number of pixels,
the typical value of an off-diagonal element of the matrix will be quite
small, so the time-ordered approach, which scales like $N_t$, will be faster
than scanning the sparse matrix, which scales like $N_{off}$.  In addition,
the RAM required by the time-ordered technique scales like $N_{pix}$ while
storing the sparse matrix scales like $N_{off}$.  Since new experiments with
large maps will have data rates only somewhat larger than \COBE's, and since
the speed of computers has grown by orders of magnitude, producing large
all-sky maps from differential data is easily feasible for future experiments.

The iterative scheme in Equation \ref{eq.iterate} can be implemented in time
order using the following scheme
\be
X^{(n+1)}_i = \frac
{\sum_t \left(\delta_{i,p(t)}\left[X^{(n)}_{m(t)}+S(t)\right]
      + \delta_{i,m(t)}\left[X^{(n)}_{p(t)}-S(t)\right]\right)}
{\sum_t \left(\delta_{i,p(t)}+\delta_{i,m(t)}\right)}
\label{eq.timeorder}
\ee
The denominator in Equation \ref{eq.timeorder} is just the diagonal of $A$,
and does not need to be computed during each iteration.  For each data point
this algorithm requires obtaining the time-ordered data $S(t), \; p(t),$ and
$m(t)$ from disk or tape, two fetches, four adds and two stores.
It goes very much faster than the actual data-taking.

\begin{figure}[t]
\plotone{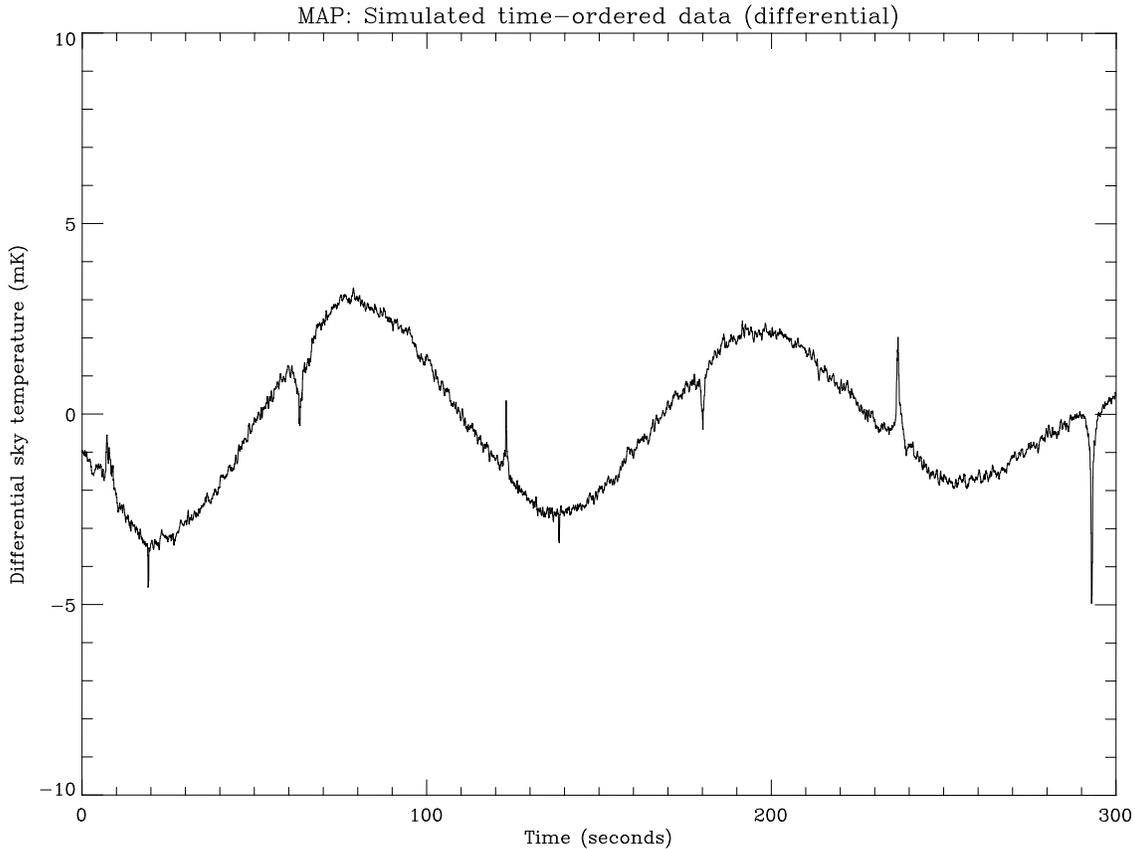}
\caption{Raw differential data from simulated space mission
observations of the sky shown in Figure \protect\ref{fig.f12f13}.
The dominant signal is the CMB dipole modulated by the motion of the
spacecraft.
Galactic plane crossings are seen as alternating spikes while the small
amplitude
fluctuations are from a cold dark matter CMB simulation.\label{fig.dtod}}
\end{figure}

\begin{figure}[t]
\plotone{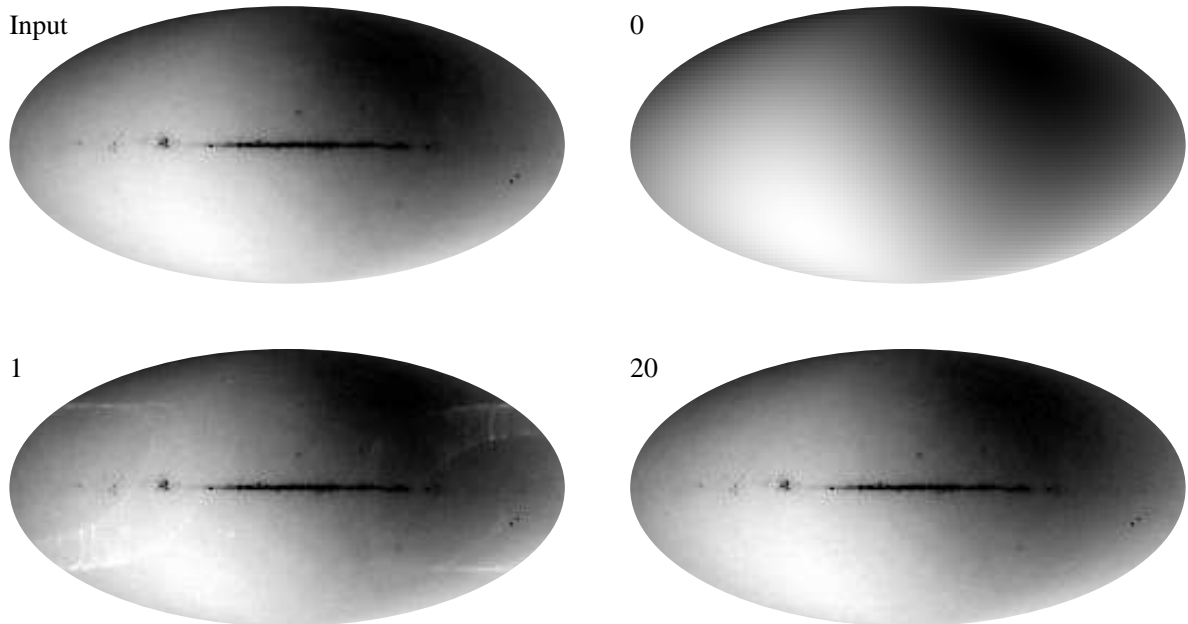}
\caption{{\it Top left}: Full sky map used as input for the
mission simulation.  The map includes simulated CMB anisotropy, the
CMB dipole, and a model Galactic signal.  {\it Top right}: The pure dipole
signal used for the $0^{th}$ iteration.  {\it Bottom left}: The recovered map
after one iteration of Equation \protect\ref{eq.timeorder}.  The Galactic
plane signal appears coherently, though with echos that are $\sim 10\%$ of
the plane signal.  This efficient reduction of plane echoes after only one
iteration requires a scan pattern that successfully connects a given sky
pixel to many other pixels, both on and off the Galactic plane.  {\it Bottom
right}: The recovered map after 20 iterations, by which time no significant
artifacts remain.\label{fig.f12f13}}
\end{figure}

In effect, this scheme evaluates the temperature in each map pixel by
averaging all the differential observations of that pixel, after correcting
each for an estimate of the signal in the reference beam, obtained from the
previous iteration.  We have tested this algorithm by simulating one year
of space mission observations using a DEC 3000/600 Alpha workstation.  The
input map, shown in Figure \ref{fig.f12f13}, contains 1,572,864 pixels and
includes both simulated CMB anisotropy, including the dipole, and a Galactic
foreground model.  We ``observed" this map with differential radiometers with
beam separation $\theta_{chop} = 135^\circ$ on a spinning spacecraft.  Figure
\ref{fig.dtod} shows a sample of the
time-ordered data from this simulation, using a data rate of 20 observations
sec$^{-1}$.  We processed one year's worth of differential observations,
including instrument noise, solving for the map using Equation
\ref{eq.timeorder}.  The iterations were started with a pure dipole map for
$X^{(0)}$.  Figure \ref{fig.f12f13} shows the input map, and the iterations
$X^{(0)}, X^{(1)}$ and $X^{(20)}$.  Note the large artifacts in the first
iteration due to the reference beam crossing the Galactic plane, which is not
modeled in $X^{(0)}$.  These artifacts are quickly removed in subsequent
iterations, and by $N=20$ there are no artifacts remaining in the map above
the level of the instrument noise.  Each iteration requires approximately 8
hours of CPU time on the existing Alpha workstation.  The required time is
proportional to the data rate and is virtually independent of the number of
map pixels: increasing the number of pixels by a factor of 4 slows processing
by only 2\%.  Significantly, 64\% of the CPU time is spent in the subroutine
that simulates the spacecraft attitude.  Thus, if applied to real data with
stored attitude information, the processing would be even faster.

This technique can be readily extended to include observations of microwave
polarization by accounting for the orientation of the observing horns with
respect to the sky.  We have tested this extension with similar mission
simulations and find suitable convergence on maps of the Stokes parameters $I$,
$Q$, and $U$.  The computational cost of including polarization is less than
a factor of 2, relative to the pure intensity case.

\begin{figure}[t]
\plotone{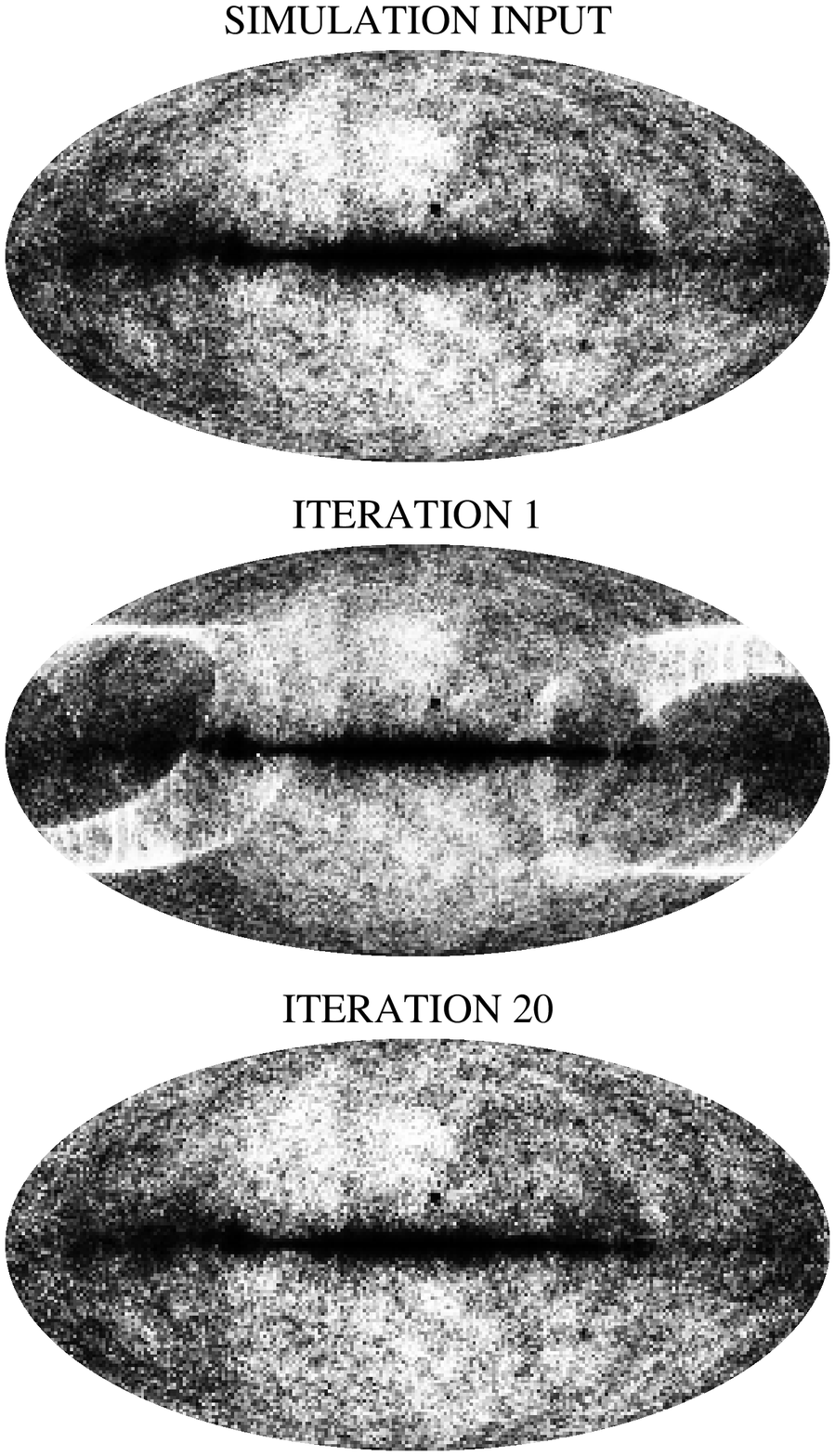}
\caption{Same as Figure \protect\ref{fig.f12f13} but with the
dipole removed and the temperature scale expanded.  The $0^{th}$ iteration is
omitted from this panel.\label{fig.ndf12f13}}
\end{figure}

\section{Discussion}

We have demonstrated a computational method that allows for the production of
mega-pixel CMB anisotropy maps from large differential radiometer data sets.
Thus the proposed new anisotropy experiments, which would produce maps with
100's of times more pixels than \COBE\ DMR's, can still use the time-tested
wide-separation differencing technique successfully employed by the DMR.

We gratefully acknowledge NASA's Office of Space Sciences for the support
provided to study the Microwave Anisotropy Probe ({\em MAP}) under the
New Mission Concepts program.

\end{document}